\documentclass{aastex}

\usepackage{amssymb}
\usepackage{graphicx}
\usepackage{emulateapj5}
\shorttitle{Isolated BCGs} \shortauthors{Zitrin et al.}

\begin{document}

%% LaTeX will automatically break titles if they run longer than
%% one line. However, you may use \\ to force a line break if
%% you desire.

\title{Star Formation Properties of Isolated Blue Compact Galaxies}

%% Use \author, \affil, and the \and command to format
%% author and affiliation information.
%% Note that \email has replaced the old \authoremail command
%% from AASTeX v4.0. You can use \email to mark an email address
%% anywhere in the paper, not just in the front matter.
%% As in the title, use \\ to force line breaks.

\author{A. Zitrin and N. Brosch}
\affil{The Wise Observatory and the Raymond and  Beverly Sackler School of Physics and
Astronomy, the Faculty of Exact Sciences, \\ Tel Aviv University, Tel Aviv 69978, Israel}
\email{adiz@wise.tau.ac.il, noah@wise.tau.ac.il}

\author{B. Bilenko}
\affil{Orbotech, Ltd.}
\email{bennybil@gmail.com}

%Modified by AZ and NB following referee remarks

\begin{abstract}
We report H$\alpha$ observations of a sample of very isolated %and void
blue compact galaxies (BCGs) located in the direction of large cosmic voids obtained to understand their stellar population compositions, the present star formation (SF) properties, and their star formation histories (SFHs). Our observations were combined with photometric data from the Sloan Digital Sky Survey (SDSS) and near-infrared data from the Two Micron All Sky Survey (2MASS), wherever such data were available. The combined data sets were compared with predictions of evolutionary synthesis models by Bruzual \& Charlot (2003a, 2003b). Current star formation rates (SFRs) were determined from the H$\alpha$ measurements, and simplified star formation histories were derived from broad-band and H$\alpha$ photometry and comparisons with the models.

We found that the star formation rates range within 0.1--1.0 M$_{\odot}$ yr$^{-1}$, with a median rate of 0.6 M$_{\odot}$ yr$^{-1}$. The observed galaxy colours are better explained by the combination of a continuous SF process with a recent instantaneous SF burst, than by a combination of several instantaneous bursts, as has been suggested previously. %In addition, the detection of current star formation in these very isolated galaxies implies the presence of material in the voids to be used for forming stars.

We compare our results for the star formation rate of the sample galaxies with that of samples of dwarf galaxies in the Virgo cluster and find that the BCGs have significantly stronger SFRs. The BCGs follow the correlation between H$\alpha$ emission and starlight found for dwarf galaxies in the Virgo Cluster and for other BCGs. %We point out that despite our initial selection of these galaxies as being extremely isolated, we found that many have a few possible companions.

\end{abstract}

%% Keywords should appear after the \end{abstract} command. The uncommented
%% example has been keyed in ApJ style. See the instructions to authors
%% for the journal to which you are submitting your paper to determine
%% what keyword punctuation is appropriate.

\keywords{galaxies: starburst; galaxies: formation; galaxies: evolution; galaxies: peculiar}

%% From the front matter, we move on to the body of the paper.
%% In the first two sections, notice the use of the natbib \citep
%% and \citet commands to identify citations.  The citations are
%% tied to the reference list via symbolic KEYs. The KEY corresponds
%% to the KEY in the \bibitem in the reference list below. We have
%% chosen the first three characters of the first author's name plus
%% the last two numeral of the year of publication as our KEY for
%% each reference.

%% Authors who wish to have the most important objects in their paper
%% linked in the electronic edition to a data center may do so by tagging
%% their objects with \objectname{} or \object{}.  Each macro takes the
%% object name as its required argument. The optional, square-bracket
%% argument should be used in cases where the data center identification
%% differs from what is to be printed in the paper.  The text appearing
%% in curly braces is what will appear in print in the published paper.
%% If the object name is recognized by the data centers, it will be linked
%% in the electronic edition to the object data available at the data centers
%%
%% Note that for sources with brackets in their names, e.g. [WEG2004] 14h-090,
%% the brackets must be escaped with backslashes when used in the first
%% square-bracket argument, for instance, \object[\[WEG2004\] 14h-090]{90}).
%%  Otherwise, LaTeX will issue an error.

\section{Introduction}
Galaxy evolution is a fundamental process in the
Universe. It includes the star formation (SF) mechanisms and accounts for  evolutionary
histories of single galaxies as well as for overall environmental effects of
galaxies on the structure and evolution of the Universe. The global
cosmological structure, and the SF processes in individual galaxies, affect each
other. Stellar evolution in galaxies enriches the heavy
element abundances and modifies the radiation field, affecting future SF in the same galaxy and its structure. Even though the SF process is fundamental, it and the different factors controlling it are not yet fully understood. Thus, investigating the importance of external effects on the star formation process in galaxies contributes to the knowledge about development of galaxies and their contribution to the evolution of the Universe.

%It is clear that gas is required as the building material for stars and that various morphological types of galaxies (E, S, S0, Irr) show different SF strengths.
It is well-known that the galaxy neighbourhood affects SF. Various studies pointed out the dependence of SF and of other properties on the density of galaxies in the immediate environment of a galaxy; this is known as the ``morphology-density'' relationship of Dressler (1980). However, while the evidence is clearer on the question of cluster environment influence, the SF behavior of galaxies in
low-density regions was and remains ambiguous. For example, Hashimoto et al. (1998) showed that the SF-environment relation is rather complicated; at similar concentration indices, the SF of galaxies in low-density regions is as strong as that of galaxies in clusters. Popescu et al. (1999) studied emission-line galaxies, specifically blue compact galaxies (BCGs) and mostly dwarf galaxies. They showed that there were no specific differences between field galaxies and objects in voids in terms of the SF rate (SFR) and the ratio between the present SFR and the past-averaged one, and that both field and void galaxies had similar SFR surface densities. Void galaxies did not exhibit lower metallicites than field galaxies, as one could expect. Gomez et al. (2003) used the Sloan Digital Sky Survey (SDSS) to study the SF-environment relation but limited themselves to field vs. cluster comparisons and to SF characterization through spectroscopy of central regions of galaxies. Haines et al. (2006) found significant differences in the star formation patterns of dwarf galaxies with --17.7$\geq$M$_R\geq$--19 as a function of neighbourhood; dwarf galaxies in the inner part of a galaxy cluster, within one virial radius, were mostly old and
passive, whereas those outside this distance and in lower density regions were forming stars.

The earliest attempt to define a sample of isolated galaxies was by Karachentseva (1973); her catalogue selection was based on distance and size of possible companion galaxies. This seelction criterion was refined in further years, including automatic sample selection from the SDSS (e.g., Allam et al. 2005).
An interest in ``isolated'' galaxies was also expressed in 1977 by Huchra \& Thuan (1977). The 12 objects selected there were observed by Brosch et al. (1982a, 1982b, 1984a, 1984b) in a range of spectral domains and the observations were summarized in a PhD thesis (Brosch 1983) with the conclusion that isolated objects show abnormal SF patterns.

The difference between isolated and other galaxies lies in the possibility that other galaxies in the vicinity of an object might trigger or influence its SF. Such influences include galaxy harassment, disruption due to tidal interactions, outright collisions, etc. At this point it is necessary to distinguish between a primary, or intrinsic, SF trigger and any number of external triggers. A primary SF trigger should not require external activation and could operate in isolated galaxies, whereas most of the time, a tidal interaction with another galaxy (especially in dwarf galaxies where these tidal effects are low) would probably not be sufficient to trigger an SF event in a relatively isolated galaxy (see also Kunth \& \"{O}stlin 2000).

Various papers studied BCGs and their SF; for example, Kong (2003, 2004, and previous papers referenced therein) investigated a large sample of BCGS with respect to their evolution histories, stellar populations, gas content and observed spectra. Pustilnik et al.\ (2001a, b) studied the influence of environment on SF in BCGs near other galaxies, and found that the SF in at least $80\%$ of them could have been triggered by tidal interactions or mergers, whereas about $43\%$ of the isolated BCGs were probably tidally influenced by dwarf galaxies in their vicinity. The influence of the environment was studied also by Balogh et al. (2004), who found higher SF in objects located in low galaxy density environments. Ceccarelli et al. (2008) found that galaxies in void walls are bluer and have more SF that galaxies elsewhere.

Pustilnik et al. (2002) investigated the HI properties of  a sample of BCG galaxies located in very low density environments. The sample comprised 26 objects from the Second Byurakan (SBS) and Case surveys located in voids with V$_{hel}\leq11000$ km s$^{-1}$, two BCGs in the void behind the Virgo cluster and 11 BCGs in denser environments. They found that at the same blue luminosity, and for M$_B\leq$-18.0 mag, BCGs in lower density environment had on average more HI.

Brosch et al. (2004) examined the possibility that the SF in dwarf irregular galaxies (DIGS) or blue compact (BCG) galaxies is triggered by tidal interactions with other nearby galaxies by searching for star-forming companion galaxies in a large (N=96) composite sample of DIGs imaged in H$\alpha$. The lack of such companions for most of the sample galaxies, along with other supporting evidence, indicates that triggering by external interactions is unlikely to be %discounted as
 a primary SF influence for these BCGs.

We aim here to study the SF properties of another sample of
galaxies known {\it a-priori} to be located in extremely low galaxy density region, by concentrating in a study of volumes located in the direction of known cosmic voids. The definition as to what exactly makes a certain volume a ``void'' is debatable since, a-priori, a void in the galaxy distribution should be completely empty of galaxies (see e.g., Lindner et al. 1999). The cosmic voids identified in studies of the large-scale structure are not completely empty of galaxies, but the object density there is extremely low, with a density contrast $\frac{\Delta \rho}{\rho} \simeq -0.94 \pm 0.02$, %i.e., close to zero galaxy density,
and showing even zero galaxy density at the centers of voids as revealed by the 2dF survey (Hoyle \& Vogeley 2004).

Hoyle et al. (2005) defined a void as a region with a galaxy
density contrast $\frac{\Delta \rho}{\rho}\leq$--0.6 over a distance scale of 7  Mpc. Using
$\sim$1000 SDSS objects, they found the void galaxies to be typically
fainter than those in the ``wall'' bordering the void. While Hoyle
et al. did not find an excess of dwarf galaxies (DGs) in the
voids, they claimed that the void population is dominated by
late-type objects. The same sample was studied spectroscopically
by Rojas et al. (2005), who found that void objects tend to have
higher specific SFRs (SFR per unit stellar mass) than wall
galaxies. Their stellar populations were also younger than for
wall galaxies.

Studies of the galaxy distribution in voids showed that the degree
of isolation is sensitive to the limiting-magnitude of the
observation (e.g., Lindner et al., 1995). The spatial distribution of
normal galaxies consists of a structure of clusters,
super-clusters, filaments and sheets of galaxies. In-between this
structure there are large voids with very low galaxy density (Lindner et al. 1996).
Lindner et al. (1997) tested the spatial distribution of BCG and
concluded that the galaxy distribution in voids is hierarchical, e.g., the
void boundaries are sensitive to the limiting magnitude of the
sample. They found that the faint galaxies in voids are concentred
toward the void boundaries, with a small fraction of the
population at the center of the voids. This fraction of faint
galaxies within the voids are grouped in systems lacking a bright
galaxy. Lindner et al. concluded that the BCG spatial distribution is similar to that of normal dwarf galaxies, with a median distance from a normal galaxy of 0.7h$^{-1}$Mpc.

Galaxies in voids were specifically targeted in the previous studies of e.g., Weistrop et al. (1995), Popescu et al. (1996), Huchtmeier et al. (1997), Popescu et al. (1999), and Grogin \& Geller (2000). Spectroscopy of 26 galaxies in the Bootes void was reported by Cruzen et al. (2002).  Zitrin \& Brosch (2008) identified a sample of nearby dwarf galaxies that trace a $\sim$0.5 Mpc linear structure in a nearby void. Interestingly, all the objects in this linear structure show low-level star formation. A possible interpretation is that the SF takes place because of the accretion of intergalactic gas onto a dark matter filament. Current $\Lambda$CDM simulations predict that low-amplitude filamentary structures criss-cross the voids (Peebles 2007).

To evaluate the influence of ``isolation'' on the SF history of a galaxy we selected a sample of galaxies that appeared to be extremely isolated and possibly located in voids. Section 2 describes the sample selection. The galaxies were observed through R-band and rest-frame H$\alpha$ filters with the Wise Observatory 1m telescope and the observations are described in \S3. The observations were supplemented by data collected from the
literature, mainly from NED\footnote{The NASA/IPAC Extragalactic
Database (NED) is operated by the Jet Propulsion Laboratory,
California Institute of Technology, under contract with the
National Aeronautics and Space Administration.}, from the Sloan
Digital Sky Survey (SDSS), and from the Two Micron All-Sky Survey
(2MASS). The analysis and results are described in \S4 along with the evolutionary models used to evaluate the star formation histories. The results are discussed in \S5 along with
a comparison with other star-forming galaxies studied by our group in a cluster environment.

\section{Sample selection}
\label{s:sample}
Different aspects must be taken into consideration in the selection of a sample of galaxies to test star forming processes. The sample selection should simplify as much as possible the
theoretical factors that affect the process. On the other hand,
 practical aspects of conducting the necessary observations,
and the amount of time the observations will require, are also
important considerations.

We chose a sample of galaxies that are relatively compact and
exhibit intense SF activity, as indicated by the presence of
strong emission lines in their spectra and by blue colours. Since
our main goal was to investigate the influence of neighbour
interaction on the process of SF, we selected galaxies that seemed to lack
known luminous objects in their neighbourhoods. Blue Compact
Galaxies (BCGs) with strong emission lines are a good selection
for this study, since a strong H$\alpha$ emission line is expected
in case of recent SF in a gas-rich object.

BCGs are known to have bluer colours [(U-B)$\sim$-0.6;
(B-V)=0.0--0.3] than ordinary irregular galaxies and their surface
brightness is much higher (Kunth 1995). Spectroscopically, in
such objects the ratio of [O III]$\lambda$5007 to H$\beta$ is
approximately 5 due to the presence of OB star complexes,
indicating a very recent burst of star formation (Kunth 1995). They %HI content of these galaxies is generally high (0.3--0.5 of the luminous mass) and their metallicity is low to very low, ranging from Z$_{\odot}$/50 to Z$_{\odot}$/3 (e.g., Izotov, Thuan \& Guseva 2005 and references therein), leading to the conclusion that these galaxies
are chemically less evolved galactic systems. Although such galaxies
were originally regarded as newly formed or primeval
galaxies, spectroscopic evidence from large samples of BCGs shows
that underneath their bright and blue stellar components lies, in most
cases, an older population of stars,
indicating previous short star formation episodes (e.g., Telles \&
Terlevich 1995; Kong 2003, 2004). Even in SBS 0335-052W, considered to be a young, recently formed galaxy, Izotov, Thuan \& Guseva (2005) found  a stellar population a few 100 Myrs old.

These properties indicate that BCG galaxies are good candidates
for our study, enabling efficient observations with a 1m telescope
even for relatively faint galaxies. The compactness of these
galaxies favours this study, since the surface brightness of the
galaxies is high and the detected signal is strong relative to
detector noise and background sky brightness.

Since our primary goal is to understand the role of the lack of frequent interactions in the regulation of the
SF, we explain first how likely are %the chances of
 interactions among very isolated galaxies.
The chance of a galaxy-galaxy interaction can be estimated from
the cross-section presented by a galaxy to a potential intruder
that could cause a star burst. In what follows we disregard the expansion of the Universe; in the past the galaxies were much closer together enhancing the interaction chance. From the arguments presented above, the impact parameter $r$ for a successful SF-triggering interaction must be 25-50 kpc.
Disregarding gravitational focusing, the number of past
interactions a galaxy experiences during a time span $\tau$ is
therefore $N\simeq \pi r^2 v n \tau$. Here $v$ is the peculiar
velocity of the target galaxy relative to its immediate galaxy
neighbourhood and $n$ is the space density of galaxies in the same
neighbourhood. This number of expected interactions during a
Hubble time can therefore be written as:
\begin{equation}
N\simeq0.05(\frac{r}{100\,kpc})^2\frac{v}{100\,km\,sec^{-1}}\frac{n}{(1\, Mpc)^3}H_{0}^{-1}
\end{equation}
with $H_{0}^{-1}$ being the age of the Universe. Given the low
velocity dispersion of galaxies in the field, and the very low
galaxy density in the vicinity of the sample galaxies, the chance of a relatively recent interaction that could have triggered SF in the objects of our sample appears very remote. Specifically, typical values of v$\simeq$200 km s$^{-1}$, r=50 kpc, and n$\leq4 \times 10^{-2}$ Mpc$^{-3}$, yield N$\leq10^{-3}$, imply a single SF-triggering interaction every 1000 Hubble times.
%Note also that a past
%interaction could have served as primary SF trigger, followed by
%SF through secondary triggers such as supernovae or stellar winds' shocks.
We also note the remark of Di Matteo et al. (2008) that strong starbursts are rare even among major galaxy
interactions or mergers.

The sample selection was aimed to minimize
the chance that an interaction with another  galaxy%, of comparable size and brightness,
 could have triggered the detected star
formation. Requiring that the distance from any other such galaxy
be larger than a given value minimizes this possibility. A
galaxy-galaxy interaction has to be fairly recent to trigger
an SF event. Icke (1985) performed numerical hydrodynamical simulations investigating the fate of a gas disk perturbed by a hyperbolic encounter with an intruder galaxy. He showed that relatively slow, pro-grade and in-plane encounters can shock the disk gas and can, in principle, cause star  formation. Struck-Marcel \& Scalo (1987) estimated
a maximal time delay between a tidal interaction and a subsequent
SF event of $\sim10^8$ years, commensurate with the rotation
period of a disk galaxy. However, the tidal triggering SF models
of Barton et al. (2000) are consistent with their observations only
for delays shorter than $\sim10^7$ years.

Galaxies that appear extremely isolated, sometimes located in central regions of voids, may really be the brighter examples of concentrations of star-forming galaxies. It is possible that SF in a seemingly isolated galaxy could be triggered by an interaction with a dwarf galaxy that is much fainter than the detection limits. One example of a dwarf irregular companion triggering SF in a void galaxy, HS 0822+3542, was suggested by Pustilnik et al. (2003). The primary object is at least 3 Mpc distant from L$_*$ galaxies but it does show signs of a recent ($\sim$10 Myr)
star burst and it has a very faint and blue companion that could be related to the starburst. Other possible neighbours, fainter than L$_*$, are dwarf galaxies at 1.15 to 1.42 Mpc
projected distance. In this case, the projected distance between
the primary galaxy and its faint companion is only 11.4 kpc and their recession
velocity difference is smaller than 25 km sec$^{-1}$. This system
is similar to the extragalactic HII regions reported by Shaver \&
Chen (1985) in that two seemingly dwarf galaxies show simultaneous
star formation bursts, and to the SBS 0335-052 system (Ekta et al. 2009).

Brosch, Bar-Or and Malka (2006) discovered small groups of uncatalogued, compact, star forming DGs by H$\alpha$ mapping of the neighbourhoods of apparently isolated, catalogued, star-forming DGs. Their sample consists of dwarf (M$_B\geq$--18 mag) galaxies at least 2 Mpc away from any other catalogued galaxy. The objects they selected as extremely isolated are conceivably the brightest members of sparse groups of dwarf galaxies that are presently forming stars. An  H$\alpha$ emission knot was identified by Brosch et al. (2007) at a projected distance of 78"$\simeq$63 kpc from the galaxy AM1934-567. Other very compact and H$\alpha$-emitting ``dots'' near other galaxies were reported by Kellar (2008) and Kellar et al. (2008), albeit generally at much closer distances from a galaxy than the objects in Brosch et al. (2006).% or the H$\alpha$ knot in AM1934-567. %Similar H$\alpha$ ``dots'' were found here and we discuss this below.
These findings imply that it is possible to find ``associated'' star-forming objects even in very sparse environments.

We constructed a sample of galaxies that a-priori meets extreme isolation criteria, based primarily on the studies of Lindner et al. (1995, 1996, 1997, 1999). The sample consists mainly of galaxies selected from the Second Byurakan Survey (SBS; Stepanian 2005 and references therein), an objective-prism emission-line galaxy survey in the sky region bordered by 7$^h$40$^m\leq\alpha\leq$17$^h$20$^m$ and 49$^{\circ}\leq\delta\leq$61$^{\circ}$, supplemented by galaxies from the Case survey (Pesch \& Sanduleak 1983 et seq., Salzer et al. 1995) in the same Right Ascension range. These objects were in the direction of identified giant voids in the galaxy distribution, and their redshifts matched those of the voids.   For these reasons we refer here to the sample galaxies as being either ``isolated'' or ``void'' galaxies.

The sample was originally chosen so that each object should have no NED objects within a projected distance $\Delta$D $\geq$3 Mpc (h=0.73) and within a redshift difference $\Delta v \geq$300 km sec$^{-1}$. The addition of redshifts to the public data sets in subsequent years (e.g., from new SDSS data releases) reduced the degree of isolation of the sample. For example, a current query of close neighbours ($\Delta$D $< 3$ Mpc, $\Delta v \leq 300$ km sec$^{-1}$, and $r \leq 25$ mag) in SDSS, or similarly in NED, reveals several neighbouring galaxies as detailed in Table \ref{sample}. Thus, some galaxies that appeared originally to be very isolated became somewhat less isolated. However, these galaxies are still located on the outskirts of galaxy filaments in the bright galaxy distribution, as can be seen when plotting the location of each object on the wedge diagrams in Fig. 3 of Lindner et al. (1996). % (Pustilnik 1997).
%{\bf Add info from SBSII}

Our decision to choose our sample primarily from the SBS and the
Case (CG) catalogs ensures that the galaxies are blue and compact,
but have strong emission lines, since they were originally detected in
objective-prism surveys. A few objects are also IRAS or FIRST sources, and many have been
observed in SDSS or other programs.

The sample galaxies are shown in Table \ref{sample}, which lists a running
number for each sample object, its leading designation, and its J2000
coordinates and radial velocity from NED. Since NED is a dynamic compilation, changes can occur at any time; one of our objects (SBS1229+579) was included in the original
sample with a redshift of 7360 km sec$^{-1}$ and H$\alpha$ observations were
performed accordingly. Later, the redshift was revised in NED to
16537 km/sec; this is way outside the range covered by the
H$\alpha$ filter used for this observation, and in fact out of the
range of any of the Wise Observatory (WO) H$\alpha$ filters, and explains why we did not detect H$\alpha$ line emission from this object. The table also lists N$_{NN}$, the number of NED companions found in the search volume (3 Mpc with H$_0$=73 km sec$^{-1}$ and 300 km sec$^{-1}$, as listed in the last checking of NED in December 2008). This indicates that the objects are in regions where the galaxy density changes from 8.8$\times$10$^{-3}$ Mpc$^{-3}$ to 0.13 Mpc$^{-3}$ and using the estimate for the number of encounters during a Hubble time from eq. 1, shows that these objects have been, in fact,  very isolated indeed with very low chances of having suffered an interaction.

%Table 2 is a cross identifier for our galaxies and contains the
%galaxy's morphological type from NED (ref.) This table also shows
%the Oxygen-Hydrogen abundance ratio

\begin{table*}
        \begin{minipage}{160mm}
\begin{center}
\caption{Primary sample of BCGs in voids} \label{sample}{\small
\begin{tabular}{rlcccl}%[SDSS ugriz + errors]
\hline No. & Name & $\alpha$, $\delta$ (J2000) & cz (km s$^{-1})$ & N$_{NN}$ & Note \\
\hline

*1a & SBS0745+601a & 07 50 01.19 +60 00 56.6 & 9773 &  &   \\

*1b & SBS0745+601b & 07 50 07.80 +60 02 26.0 & unknown &  & Blue, starlike object \\

*2 & SBS0749+582 & 07 53 49.99 +58 09 09.5 & 9548 &  & \\

*3 & SBS0805+607 & 08 09 30.81 +60 36 46.3 & 9384 &  &  \\

4 & SBS0813+521 & 08 17 36.99 +52 02 35.6 & 7052 & 2 &  Two H$\alpha$ regions \\
%17.171 16.379
%16.214 16.175 16.045 0.012 3.472E-3 3.793E-3 4.211E-3 0.011
%no SDSS spectrum\\

5 & G0834+3614  & 08 37 12.73 +36 04 06.7 & 9920 & 0 & \\
%18.245 17.445 17.177 17.019 16.935 0.02 0.023 7.694E-3 9.004E-3 0.021 \\

6 & G0919+3626  & 09 22 31.17 +36 13 33.7 & 9218 & 2 & \\
%17.33 16.612 16.347 16.171 16.021 0.011 3.778E-3 3.933E-3 4.772E-3 0.011 \\

7 & SBS0938+611 & 09 42 36.52 +60 52 34.3 & 7969 & 6 & MRK 1421 \\
%17.483 16.478 15.959 15.697 15.462 0.014 4.853E-3 4.193E-3 4.725E-3 7.888E-3 \\

8 & G0942+3645  & 09 45 42.50 +36 31 26.8 & 9876 & 1 & \\
%16.976 15.984 15.663 15.441 15.318 8.719E-3 2.937E-3 2.872E-3 3.185E-3 6.003E-3 \\

9 & G0957+3503  & 10 00 18.29 +34 48 52.8 & 11700 & 15 & CG 0313 \\
%18.567 17.69 17.46 17.248 17.136 0.029 6.264E-3 7.065E-3 8.757E-3 0.029
%no SDSS spectrum\\

10 & SBS1032+496 & 10 35 08.52 +49 21 41.3 & 8553 & 4 & \\
%17.787 17.138 16.8 16.636 16.505 0.014 7.01E-3 6.377E-3 7.441E-3 0.014 \\

11 & SBS1118+587 & 11 21 35.70 +58 29 27.1 & 8493 & 5 & \\
%18.309 17.557 17.495 17.486 17.407 0.023 5.899E-3 6.804E-3 8.078E-3 0.024 \\

12a & SBS1120+586 & 11 23 39.64 +58 22 42.6 & 11097 & 1 & Galaxy Pair w. 12b\\
%18.872 18.204 18.127 17.994 18.007 0.025 9.301E-3 9.329E-3 0.011 0.031 \\
12b & SBS1120+586B & 11 23 48.3 +58 22 05.00 & 11198 & 1 & Galaxy Pair w. 12a \\

13 & SBS1122+610 & 11 25 14.64 +60 46 58.5 & 9787 & 6 & \\
%18.254 17.594 17.517 17.385 17.356 0.016 5.973E-3 6.356E-3 0.01 0.021 \\

14 & SBS1221+602 & 12 23 22.53 +59 56 29.7 & 7074 & 1 & \\
%17.272 16.507 16.246 16.085 15.899 0.013 3.614E-3 3.693E-3 4.283E-3 0.011 \\

15 & SBS1229+578 & 12 31 43.72 +57 32 07.9 & 16537 & & Rejected from the sample \\
%17.56 16.551 16.223 15.99 15.872 0.022 3.903E-3 4.174E-3 4.674E-3 0.016
% cz as listed, Sab; S Ha Hb OII\\

16 & SBS1235+559 & 12 37 36.17 +55 41 03.7 & 8785 & 0 & \\
%18.637 17.709 17.647 17.554 17.626 0.04 7.572E-3 9.332E-3 0.014 0.053 \\
17a & SBS1311+563a & 13 13 15.9 +56 05 52.00 & 11970 & 3 & MRK 0246, Galaxy Pair\\

17b & SBS1311+563b & 13 13 56.47 +56 07 37.4 & 12006 & 3 & Galaxy Pair\\
%18.1 17.316 17.227 17.046 16.989 0.025 6.202E-3 6.912E-3 9.262E-3 0.022 \\

18 & SBS1323+575 & 13 25 52.96 +57 15 15.1 & 6287 & 5 & MRK 66 \\
%15.962 15.006 14.811 14.695 14.675 5.682E-3 2.324E-3 2.333E-3 2.531E-3 4.502E-3 \\

19 & SBS1332+599 & 13 34 43.94 +59 44 33.9 & 9041 & 5 & \\
%17.368 16.421 16.182 16.048 15.906 0.017 3.86E-3 4.01E-3 4.978E-3 0.013 \\

20 & SBS1353+597 & 13 55 35.03 +59 30 42.1 & 6713 & 1 & \\
%17.856 16.824 16.463 16.283 16.107 0.025 7.839E-3 5.804E-3 6.6E-3 0.024
%edge-on spiral?\\

21 & SBS1354+580 & 13 56 23.35 +57 45 44.7 & 8393 & 0 & \\
%17.347 16.417 16.165 15.986 15.791 0.015 3.578E-3 3.869E-3 4.424E-3 0.013
%v. nearby star/galaxy?\\

22 & SBS1408+558 & 14 09 46.50 +55 35 57.5 & 7781 & 1 &  KUG 1408+558 \\

*23 & SBS1420+544 & 14 22 38.74 +54 14 07.9 & 6176 & \\

*24 & G1427+3343  & 14 29 31.96 +33 30 35.0 & 7910 & \\

*25 & G1430+3316  & 14 32 19.01 +33 02 48.4 & 11055 & \\

26 & G1445+3801  & 14 47 35.18 +37 49 01.3 & 10462 & 0 & \\

27 & SBS1457+540 & 14 58 40.44 +53 51 29.2 & 7867 & 9 & \\
%17.959 16.387 15.639 15.263 14.855 0.026 3.85E-3 3.309E-3 3.686E-3 7.035E-3
%interacting/overlapping pair; no SDSS spectrum\\

28 & SBS1527+583 & 15 28 38.53 +58 12 43.4 & 5814 & 0 & \\

29 & SBS1541+515 & 15 43 02.47 +51 25 47.7 & 10577 & 0 & \\
%18.543 17.463 17.362 17.215 17.262 0.027 5.557E-3 6.316E-3 8.732E-3 0.028
%no SDSS spectrum\\

*30 & SBS1556+583 & 15 57 54.97 +58 09 39.3 & 10360 & & MRK 0865\\
 \hline

\end{tabular}
}
\end{center}
%\note{
Notes: Objects marked with an asterisk have no SDSS nor WO photometry; although selected and listed here, they were not observed in H$\alpha$ and are not included in the following analysis. Object no.\ 15 has had its NED redshift updated to a higher value; this excludes it from the redshift range of our sample.
    \end{minipage}
\end{table*}

We note that a large number of objects in our sample, originally catalogued as single objects, were commented on as being very close pairs of galaxies. Stepanian (2005) mentions this for no less than eight objects listed in Table~1 (e.g., SBS0745+601 with the mention ``close binary with angular separation $\sim$3 arcsec in a common spheroidal shell''). Note that in such cases it may be possible to interpret the two bright condensations in a single envelope as parts of a single galaxy, instead of assuming a very close pair. For this reason we retain all the eight cases where Stepanian mentions binarity in the sample, though we quote his comments below. This is true for other objects in Table 1, where similar remarks were made by Pustilnik et al. (2002), or a possible binarity is visible by just inspecting the SDSS images.

\section{Observations and data reduction}

Optical imaging photometry observations were carried out at the
Wise Observatory in Israel with the 1.0-m telescope, mainly during 1996-2000 (hereafter called ``primary observations''), and were supplemented by additional observations in the beginning of 2008.

%\subsection{Broad-band imaging}
Each sample galaxy was imaged through a standard R filter with a total exposure time of 30 to 60 min, and through an appropriate rest-frame $H\alpha$ narrow filter for a total exposure time of 60-100 min. All the primary observations were done with a cryogenically-cooled 1024x1024 pixel Texas Instruments CCD with 12 $\mu$m square pixels $\simeq$0.4" covering $\sim5'\times5'$ at the f/7 Ritchey-Chr\'{e}tien focus. The 2008 observations were done with a Princeton Instruments VersArray camera equipped with a $1340\times1300$ pixel CCD chip with a plate scale of 0.58"/pixel and a $12.95'\times12.57'$ field of view.

Each observation was split between two to four images per filter, so that cosmic rays and hot pixels could be detected and removed at the image reduction stage. R-band images of the sample galaxies with SDSS photometry were calibrated with the $r$ magnitudes of reference stars in the galaxy fields, calculated from the SDSS photometry as derived by Lupton in 2005 and given in the SDSS DR6 web page\footnote{http://www.sdss.org/dr6/algorithms/sdssUBVRITransform.html}.

%\subsection{ measurements}
The sample galaxies were observed through narrow-band FWHM$\sim$50\AA\ H$\alpha$ filters
using the same CCD as for the broad-band imaging to determine
their line emission. Each galaxy was imaged through
a filter approximately centered on the rest wavelength of the
line. The central wavelengths of the H$\alpha$
filters range from 6630\AA\, to 6800\AA\,; the filters cover the
velocity range of the galaxies, but for some galaxies the
central wavelength may be shifted by up to 20\AA\, from the
expected rest-frame H$\alpha$ line, given the published redshift.
This effect is taken into account in the derivation of the H$\alpha$ line flux and its error. However, we did not correct the measurements for possible contamination by [NII] emission.
%The net  Under the assumption that stars do not contribute
%significantly to the galaxy H$\alpha$ line (given their zero
%redshift) the stars will not contribute radiation in the line, and
%thus can be used to determine the continuum emission in the
%galaxies.

%We used some of the WO H$\alpha$ filters listed in Table~\ref{t:filters}, which were re-measured in summer 2006. The table lists the central wavelength of the filter $\lambda_c$ and  the full-width at half-transmission of the filter $\Delta \lambda$. We also list the redshift ranges (cz), over which the filters can be used. These include all velocities for which the shifted H$\alpha$ line would be within  0.3$\times$FWHM (Full Width at Half Maximum) of the filter's Central Wavelength $\lambda_C$.  %Note that even though the peak-to-peak ``redshift distance'' between adjacent filters is $\sim 15h^{-1}$ Mpc, the significant transmission at the half-power point of the transmission profiles ensures that there will not be many redshift ranges lacking coverage.

%As these are interference filters, the transmitted bandpass depends on the incident angle of the incoming radiation. The influence of having the narrow-band filter near the focal plane of the f/7 in the  converging beam of the telescope has been tested by Almoznino (1996 PhD thesis). The central wavelength shifts because of angle-of-incidence effects by 1.5-3.3\AA\, to the blue; this is a negligible shift for our purposes, considering the transmission profile of the filter.

H$\alpha$ images of these galaxies were calibrated using the effective ratio of the H$\alpha$ and R filters, assuming that the reference stars do not have significant H$\alpha$ emission or absorption. For all galaxies, the R filter was used for continuum measurement of the $H\alpha$ line. The net emission from the H$\alpha$ line was calculated by normalizing the H$\alpha$ and the R images using reference stars in the field and correcting for the filter transmission.

Each H$\alpha$ exposure collected for this program was 1200 sec
long. This duration is a compromise between cosmic ray
contamination of the CCD frame and imaging depth; by combining a
number of exposures we can confidently
detect individual HII regions in Virgo cluster (VC) dwarf irregular galaxies (F$_{H\alpha} \geq 10^{-15}$ erg cm$^{-2}$ sec$^{-1}$, Heller et al. 1999)
and eliminate most cosmic ray events. The targets observed here
are between two to five times more distant than the VC but are, in
most cases, resolved showing non-stellar images. %Otherwise, their intrinsic H$\alpha$ flux should have been stronger by a factor of 4-25 than that of the VC HII regions we studied in order to reach the quoted detection limits. The use of i
Interference filters have one drawback; one sometimes sees reflection ghosts in the
vicinity of bright objects. To overcome this
problem, at least three slightly dithered exposures were collected
for each field, representing one hour or more of telescope time.
This combination allowed the rejection of most ghost reflections
and cosmic rays in the final image.

%\subsection{Photometric calibration}

%\section{}
The image processing included standard bias subtraction,
flat-fielding, and sky subtraction. Individual images obtained
through the same filter were aligned, cropped to the same size,
and combined by a median filter into a final filter image. The H$\alpha$ fluxes measured in these final images were calibrated using the effective widths of the H$\alpha$ and continuum (R band) filter ratios.

%subsection{A data} \label{MoreData}

Additional broad band photometric data were collected from SDSS and 2MASS. The {\it ugriz} photometry from SDSS  and the JHK photometry from 2MASS are presented in Tables \ref{SDSS} and \ref{2mass}, respectively. The entries contain also the measurement error (in parenthesis). %As mentioned, galaxies 1-3 had no SDSS data and our UBVRI photometry is presented instead. W
The absolute B magnitudes of the galaxies are listed in column 2 of Table 4; we note that all the objects are fainter that M$_B$=--20 mag; while about 2/3 of the objects are not dwarf galaxies (M$_B\leq$--18 mag), they are nevertheless fainter than L$_*$. Below we  add  short individual descriptions from Stepanian (2005), Pustilnik et al. (2002), and others for some of the objects.

\begin{table*}[h]

%wide two-column table
        \begin{minipage}{160mm}
\begin{center}
\caption{UBVRI and SDSS photometry of the sample galaxies}\label{SDSS} {\small
\begin{tabular}{cccccc}
\hline

No. & u & g & r & i & z \\ \hline

4 & 17.17(0.01) & 16.38(0.01) & 16.21(0.01) & 16.18(0.01) & 16.05(0.01) \\

5 & 18.25(0.02) & 17.45(0.02) & 17.18(0.01) & 17.02(0.01) & 16.94(0.02) \\

6 & 17.33(0.01) & 16.61(0.01) & 16.35(0.01) & 16.17(0.01) & 16.02(0.01) \\

7 & 17.48(0.01) & 16.48(0.01) & 15.96(0.01) & 15.70(0.01) & 15.46(0.01) \\

8 & 16.98(0.01) & 15.98(0.01) & 15.66(0.01) & 15.44(0.01) & 15.32(0.01) \\

9 & 18.57(0.01) & 17.69(0.01) & 17.46(0.01) & 17.25(0.01) & 17.14(0.01) \\

10 & 17.79(0.01) & 17.14(0.01) & 16.80(0.01) & 16.64(0.01) & 16.51(0.01) \\

11 & 18.31(0.02) & 17.56(0.01) & 17.50(0.01) & 17.49(0.01) & 17.41(0.02) \\

12a & 18.87(0.03) & 18.20(0.01) & 18.13(0.01) & 17.99(0.01) & 18.01(0.03) \\

12b &18.83(0.03) & 17.98(0.01) & 17.69(0.01) & 17.47(0.01) & 17.30(0.02)\\

13 & 18.25(0.02) & 17.59(0.01) & 17.52(0.01) & 17.38(0.01) & 17.36(0.02) \\

14 & 17.27(0.01) & 16.51(0.01) & 16.25(0.01) & 16.09(0.01) & 15.90(0.01) \\

15 & 17.56(0.02) & 16.55(0.01) & 16.22(0.01) & 15.99(0.01) & 15.87(0.02) \\

16 & 18.64(0.04) & 17.71(0.01) & 17.65(0.01) & 17.55(0.01) & 17.63(0.05) \\

17a &  17.07(0.01) & 16.06(0.01) & 15.79(0.01) & 15.54(0.01) & 15.40(0.01) \\

17b & 18.10(0.03) & 17.32(0.01) & 17.23(0.01) & 17.05(0.01) & 16.99(0.02) \\

18 & 15.96(0.01) & 15.01(0.01) & 14.81(0.01) & 14.70(0.01) & 14.68(0.01) \\

19 & 17.37(0.02) & 16.42(0.01) & 16.18(0.01) & 16.05(0.01) & 15.91(0.01) \\

20 & 17.86(0.03) & 16.82(0.01) & 16.46(0.01) & 16.28(0.01) & 16.11(0.02) \\

21 & 17.35(0.02) & 16.42(0.01) & 16.17(0.01) & 15.99(0.01) & 15.79(0.01) \\

22 & 16.94(0.02) & 15.67(0.01) & 15.21(0.01) & 14.94(0.01) & 14.74(0.01) \\

%23 & SBS1420+544 & 14 22 38.74 +54 14 07.9 & & \\

%24 & G1427+3343  & 14 29 31.96 +33 30 35.0 & & \\

%25 & G1430+3316  & 14 32 19.01 +33 02 48.4 & & \\

26 & 17.47(0.01) & 16.84(0.01) & 16.65(0.01) & 16.45(0.01) & 16.32(0.01) \\

27 & 17.96(0.03) & 16.39(0.01) & 15.64(0.01) & 15.26(0.01) & 14.86(0.01) \\

28 & 17.82(0.02) & 17.04(0.01) & 16.77(0.01) & 16.62(0.01) & 16.56(0.03) \\

29 & 18.54(0.03) & 17.46(0.01) & 17.36(0.01) & 17.22(0.01) & 17.26(0.03) \\

 \hline

\end{tabular}
}
\end{center}
    \end{minipage}
\end{table*}

%we list in Table~\ref{t:2mass} the total H, J, and K magnitudes from 2MASSS, with the photometric errors in parentheses.

\begin{table*}[h]
%wide two-column table

        \begin{minipage}{160mm}
\begin{center}
\caption{2MASS photometry %, and SDSS spectroscopy
of the sample galaxies} \label{2mass}{\small
\begin{tabular}{rccccccc}
\hline
%\tablehead{\colhead{Date} & \colhead{Camera 1 yield}   &
%\colhead{Camera 2 yield} } \startdata
No. &  J$_T$ & H$_T$ &K$_T$\\ \hline

4 & 15.08(0.16) & 14.55(0.25) &  13.99(0.22) \\

%5 & 18.245(0.02) & 17.445(0.023) & 17.177(7.694E-3) & 17.019(9.004E-3) & 16.935(0.021) \\

6 & 14.80(0.13) & 14.29(0.20) & 13.76(0.20 \\

7 & 14.22(0.08) & 13.69(0.11)  & 13.31(0.14)  \\

8 & 14.41(0.08)   & 13.80(0.12) & 13.31(0.12)   \\

%9 & 18.567(0.029) & 17.69(6.264E-3) & 17.46(7.065E-3) & 17.248(8.757E-3) & 17.136(0.029) \\

%10 & 17.787(0.014) & 17.138(7.01E-3) & 16.8(6.377E-3) & 16.636(7.441E-3) & 16.505(0.014) \\

%11 & 18.309(0.023) & 17.557(5.899E-3) & 17.495(6.804E-3) & 17.486(8.078E-3) & 17.407(0.024) \\

%12 & 18.872(0.025) & 18.204(9.301E-3) & 18.127(9.329E-3) & 17.994(0.011) & 18.007(0.031) \\

%13 & 18.254(0.016) & 17.594(5.973E-3) & 17.517(6.356E-3) & 17.385(0.01) & 17.356(0.021) \\

14 & 14.81(0.15) & 14.48(0.27) & 13.90(0.28) \\

15 & 15.03(0.16) & 14.37(0.22) & 13.71(0.19) \\

%16 & 18.637(0.04) & 17.709(7.572E-3) & 17.647(9.332E-3) & 17.554(0.014) & 17.626(0.053) \\

%17 & 18.1(0.025) & 17.316(6.202E-3) & 17.227(6.912E-3) & 17.046(9.262E-3) & 16.989(0.022) \\

18 & 13.57(0.05) & 13.18(0.08) & 12.91(0.11) \\

19 & 15.26(0.23) & 14.48(0.28) & 13.84(0.26) \\

%20 & 17.856(0.025) & 16.824(7.839E-3) & 16.463(5.804E-3) & 16.283(6.6E-3) & 16.107(0.024) \\

%21 & 17.347(0.015) & 16.417(3.578E-3) & 16.165(3.869E-3) & 15.986(4.424E-3) & 15.791(0.013) \\

22 & 13.95(0.06) & 13.11(0.07) & 13.19(0.13) \\

%23 & SBS1420+544 & 14 22 38.74 +54 14 07.9 & & \\

%24 & G1427+3343  & 14 29 31.96 +33 30 35.0 & & \\

%25 & G1430+3316  & 14 32 19.01 +33 02 48.4 & & \\

26 & 15.11(0.16)  & 14.59(0.24) & 14.06(0.26) \\

27 & 13.84(0.06) & 13.20(0.09) & 12.91(0.10)\\

%28 & SBS1527+583 & 15 28 38.53 +58 12 43.4 & & \\

%29 & 18.543(0.027) & 17.463(5.557E-3) & 17.362(6.316E-3) & 17.215(8.732E-3) & 17.262(0.028) \\

30 & 13.86(0.06) & 13.16(0.08) & 12.82(0.09)  \\

 \hline

\end{tabular}
}
\end{center}
    \end{minipage}
\end{table*}

%{\bf ADD data from the Case survey and NED comments}

{\bf SBS0745+601} is described by Stepanian (2005) as a close binary galaxy with two star-like components in a common shell. He mentions that this object shows strong H$\alpha$, H$\beta$ and [OIII] emission are seen from the two components.

{\bf SBS0749+582} is a starlike BCDG, with a weak continuum that did not record on the SBS objective-prism plates (Stepanian 2005). He II 4686\AA\ emission is present.

{\bf G0834+3614} has a nearby H$\alpha$-emitting galaxy identified in SDSS as J083723.38+360448.1. With a radial velocity of 9108 km sec$^{-1}$, this object should not be accepted as a true companion, but presumably represents a foreground object.

{\bf G0919+3626} is noted by Pustilnik et al. (2002) as elongated and with a disturbed NE part. A faint possible companion is present 15" to the East. The SDSS image shows that this proposed companion might be part of a blue polar ring.

{\bf SBS1032+496} was noted by Pustilnik et al. (2002) as having a disturbed external morphology.

{\bf SBS1120+586} is listed by Stepanian (2005) as two separate objects; ``A'' and ``B'', with the A component carrying the mention that it is a ``close-binary'' and B being a ``blue strip of knots, elongated along right ascension''. Pustilnik et al. (2002) noted the small projected separation (48 kpc) and velocity difference (15 km sec$^{-1}$).

{\bf SBS1122+610} was classified as BCDG by Stepanian (2005), who noted a very well formed starlike nucleus seen on the slit of the prime focus of the 6m BTA telescope.

{\bf SBS1221+602} was noted by Pustilnik et al. (2002) as having a probable faint galaxy in contact 21" to the West.

{\bf SBS1311+563} is a pair consisting of MRK 254 and a companion 5'.8 away.

{\bf SBS1332+599} was noted by Pustilnik et al. (2002) as having faint galaxy in contact 15" to the South.

{\bf SBS1353+597} has  a faint r=17.89 companion $\sim$15" to the SW.

{\bf SBS1354+580} noted as a close binary galaxy, or as having a B$\simeq$18.5 companion $\sim$5 arcsec away (Stepanian 2005). Pustilnik et al. (2002) qualified it as an advanced merger, whereas SDSS shows that it is a contact system of a bluish disk and an elliptical.

{\bf SBS1420+544:} Pustilnik et al. (2002) noted that it is a large disk with several possible companions within less than 40".

{\bf SBS1420+544} is a BCDG according to Stepanian (2005). The objective prism spectra do not show a trace of the continuum.

{\bf G1457+540} shows a number of possible contact companions on the SDSS image. Lacking redshift information for these precludes a determination whether they are only projected companions.

{\bf SBS1457+540} was noted as a close binary by Stepanian (2005). The SDSS image shows that is a contact system of a bluish disk with an elliptical.

{\bf SBS1541+515} was noted as a close binary with a separation of $\sim$5 arcsec (Stepanian 2005).

Only four of the SBS galaxies in our sample were detected by IRAS and only two are FIRST radio sources.  In general, we note the relatively large fraction of objects classified as close binary galaxies, as the comments on individual objects show. However, we also stress that at present there is no redshift confirmation that these putative neighbours are real and not just seen in projection.

\section{Analysis and results}

%% The displaymath environment will produce the same sort of equation as
%% the equation environment, except that the equation will not be numbered
%% by LaTeX.
We describe below the derivation of the current star formation rate and the simplified star formation histories of the sample galaxies.
The SFR is derived directly from the H$\alpha$ line emission by:
\begin{equation}
SFR[M_{\odot}/yr] = 1.27\cdot10^{9}F_{H\alpha}\cdot D^{2}
\end{equation}
where $D$ is in Mpc and was calculated assuming Hubble expansion with $H_{0}=73$\ km/s/Mpc, and $F_{H\alpha}$ is in erg~cm$^{-2}$~s$^{-1}$. As mentioned above, the H$\alpha$ flux was calibrated by the effective ratio of the H$\alpha$ and R filters, using the R magnitudes of the secondary stars calculated from the SDSS magnitudes, and was corrected for the specific filter transmission profile. The transformation from the real R magnitudes to flux  used the zero point of Bessel (1979) and the central wavelength of the R filter. However, we did not correct $F_{H\alpha}$ for internal extinction; we consider such corrections very uncertain irrespective of whether they are derived from the H$\alpha$/H$\beta$ line ratio from the (central 3 arcsec) SDSS spectra or from a guesstimate of the dust content using
IRAS fluxes. For this reason, the SFR values given here should be considered lower limits.

In order to derive a simplified SFH, we used galaxy evolution models (GEMs) to produce photometric data to be compared to those presented above for the sample galaxies. The GEMs used are part of the GALAXEV library (Bruzual \& Charlot 2003a, 2003b). These were calculated using the Padova 1994 evolutionary tracks, with a Salpeter (1955) IMF and standard lower and upper mass cutoffs of $m_{L}=0.1M_{\odot}$ and $m_{U}=100M_{\odot}$. The models calculate the colours of the galaxy for 220 time steps from 20Gyr ago to the present. As mentioned above, BCGs are known to have low metallicity (Z$_{\odot}$/50 to Z$_{\odot}$/3). Therefore, and in order to lower the number of free parameters to be fitted, we used single metallically models with Z=0.2 Z$_{\odot}$. Furthermore, most integrated colours do not change significantly as a function of metallicity and, since we are interested mainly in the timing of the SF bursts (or processes), this is a fair assumption.

We created an $H\alpha$-V colour that provides another parameter to be compared with the models. This, in order to enhance the reliability of the fit and to lower the degeneracy. The models predict the number of Lyman continuum (LyC) photons, which depends on the distance to each object. To remove the uncertainty related to the lack of accurate distances to all of the galaxies, we used the prediction to derive a distance-independent colour for each galaxy to which the observed colour can be compared. The $H\alpha$ luminosity is related to the flux of LyC photons by:% (Osterbrock 1989):
\begin{equation}
N_{c} = 7.43 \times 10^{11}L(H\alpha)    ,
\end{equation} \newline
where $L(H\alpha)$ is in erg/s. This leads to:
\begin{equation}
H\alpha-V = 129.8 -2.5 \times log[N_{c}] - V_{abs} ,
\end{equation} \newline
where both $log[N_{c}]$ and $V_{abs}$ [mag] are given by the model. Since for most  galaxies we use SDSS photometry, their V magnitudes were calculated using the equations derived by Lupton from the SDSS DR6 web page. For more information on the model comparison procedure see Zitrin \& Brosch (2008).

The default models describe the colours of a single generation of stars formed in an instantaneous SF burst at different time steps, while the real stellar population might be a mix of a number of SF processes or burst-formed populations. Therefore a script was written to find by $\chi^{2}$ the best combinations of colours and burst times from the data base. Our models first used a combination of three bursts; this reveals, contrary to models including only one or two bursts, the old population underlying the newly formed stars.  Such a significant population of old stars was also found in other BCGs (e.g., Kong et al. 20003).

Subsequently, we found that a combination of a constant SF process underlying a recent SF burst describes the galaxy colours better than a combination of 2-3 short bursts at odd intervals, as it reduces the typical $\chi^{2}$ on average by an order of magnitude. We therefore created, using the GALAXEV program, models with a constant SFR (0.1--1.0 M$_{\odot}$ yr$^{-1}$) and used its combination with the instantaneous burst models to compare to the observed colours. All available colours for each galaxy were used in the comparisons with the models. The script outputs the 10 best fits. The dispersion of these nine additional results around the best fit, in terms of burst times and their weights, was used as the error of the comparison. The dispersion around the best fit was typically a few Myr for the recent bursts and about $\sim1$Gyr for the ``old'' bursts (or the beginning of the constant SF process). The dispersion of the relative weights of the bursts (the fraction of the light measured nowadays) was $\sim10\%$ around the best-fit values. In addition, since the time resolution is low for early bursts, very old events are only indicated as taking place more than 10 Gyr ago, without an attempt to specify the exact time.

The $\chi^{2}$ statistic calculated for each object is:
\begin{equation}
\chi^{2} = \frac{\sum \frac{(\Delta colour)^{2}}{err^{2}}}{N(col)-N(burst)}.
\end{equation}
where $\Delta colour$ is the difference between each measured colour and the linearly-combined model-produced colour, $err$ is the error of the measured colour, $N(col)$ is the number of colours used for the fit and $N(burst)$ is the number of star formation bursts modeling the SFH, and the sum refers to the colours used for the fit. The script selects the minimal value of $\chi^2$ that indicates the best-fitted model.

For galaxies lacking 2MASS data, the measurable quantities used for the fit are only the five colours: (H$\alpha-g), (u-g), (g-r), (r-i), (i-z)$. To those we fit the fraction of measured light produced by stars formed in a specific burst, and the time of the burst. For two events (single instantaneous burst combined with a constant SFR) we have three free parameters: the percentage of present light produced by one SF component, the constant SFR beginning time, and the time of the instantaneous burst. This implies that the fit is exactly determined. Adding the 2MASS information introduces two additional colours; this improves the determination of the solution.

The results of the fit are shown in Table~\ref{models}. We list there the running number of the specific galaxy in our sample as given in Table~1, its absolute B magnitude, the measured total H$\alpha$ flux, the equivalent width of the H$\alpha$ line in \AA, the present star formation rate in M$_{\odot}$ yr$^{-1}$, and the fitted constant SFR and recent burst solution, where for each event we list the time of the SF start (column 6 refers to the continuous SF and column 7 to the instantaneous burst) and its contribution to the light measured at present (columns 8 and 9). The last column is the minimal $\chi^2$ of the fit defined above. The H$\alpha$ line flux errors are typically 15\% and are contributed by the measurement error, the uncertainty and assumptions of the calibration process, the filter ratio errors, and a minor uncertainty regarding the effective central wavelength.

\begin{table*}[!h]
\label{models}
\begin{minipage}{160mm}
%\begin{center}
%wide two-column table
        %\begin{minipage}{160mm}

\begin{center}
\caption{H$\alpha$ line fluxes, SFR and SFH of the sample galaxies}
%\label{bursts}
\begin{footnotesize}
\begin{tabular}{rccccccccccc}
\hline No. &  M$_{B}$ & H$\alpha$ & EW & SFR  & B1 & B2 &  \%B1 & \%B2 & $\chi^{2}$ \\
           &   (mag)  &  erg cm$^{-2}$ s$^{-1}$  & (\AA) & (M$_{\odot}$ yr$^{-1}$) & & & & & \\
\hline
4 &-18.27& 5.79$\times10^{-14}$ & 41.41& 0.69 & 1.6Gyr& 1.3Gyr& 0.87& 0.13 & 0.60 \\
5 &-17.94& 2.44$\times10^{-14}$  & 39.35&0.57 & $>10$Gyr& 7.9Myr & 0.89 & 0.11 &0.01\\
6  & -18.63&3.91$\times10^{-14}$ &30.13&0.79 & $>10$Gyr& 10.9Myr & 0.8 & 0.2 & 0.48\\
7  &-18.39& 5.77$\times10^{-14}$  &35.42&0.87 & $>10$Gyr& $>10$Gyr & 0.93 & 0.07 & 2.74\\
8   &-19.36&4.90$\times10^{-14}$  &26.04&1.14 & $>10$Gyr& 1.3Gyr & 0.84 & 0.16 & 0.49\\
9   &-18.04&9.30$\times10^{-15}$  &24.78 &0.30 & $>10$Gyr& 143Myr & 0.88 & 0.12 & 0.10\\
10  &-17.95&7.84$\times10^{-14}$  &51.48&1.36 & $>10$Gyr& 1.4Myr & 0.83 & 0.17 & 3.36\\
11  &-17.50&3.31$\times10^{-14}$  &51.30&0.57 & 1.1Gyr& 255Myr & 0.96 & 0.04 & 0.44\\
12a &-17.45&1.17$\times10^{-14}$  &39.01&0.34 & 1.6Gyr& 55Myr & 0.78 & 0.22 & 0.11\\
12b & -17.66 &1.92$\times10^{-14}$ &40.41& 0.57 & 1.7 Gyr & 29 Myr & 0.94 & 0.06 & 0.07\\
13  &-17.79& 1.58$\times10^{-14}$ &31.12& 0.36 & $>10$Gyr& 11Myr & 0.80 & 0.2 & 0.42\\
14  &-18.15&5.36$\times10^{-14}$ &39.26& 0.64 & $>10$Gyr& $>10$Gyr & 0.93 & 0.07 & 1.38\\
15  &-19.91& --- &--- & ---& $>10$Gyr& 570Myr & 0.99 &0.01 & 1.67 \\
16 &-17.39&1.19$\times10^{-14}$ &31.63& 0.22 & 1.1Gyr & 508Myr & 0.81 & 0.19 & 1.88\\
17b  &-18.49&1.52$\times10^{-14}$ &33.58& 0.52 & 3Gyr & 161Myr & 0.85 & 0.15 & 0.18\\
18 &-19.36&4.65$\times10^{-13}$ & 44.61&4.38 & $>10$Gyr& 3.3Myr & 0.78 & 0.22 & 1.21\\
19  &-18.73&3.57$\times10^{-14}$ & 36.20 &0.69 & $>10$Gyr& 1.6Myr & 0.84 & 0.16 & 4.61\\
20  &-17.67&1.97$\times10^{-14}$ & 25.21& 0.21 & $>10$Gyr& 1.3Gyr & 0.74 & 0.26 & 0.14\\
21  & -18.58&4.86$\times10^{-14}$ & 26.21&0.82 & $>10$Gyr& 360Myr & 0.88 & 0.12 & 0.31\\
22 & -19.12&3.06$\times10^{-14}$ & 23.41&0.45 & $>10$Gyr& $>10$Gyr & 0.59 & 0.41 & 0.84\\
26  &-18.70&9.51$\times10^{-14}$ & 56.69&2.48 & $>10$Gyr& 2Myr & 0.68 & 0.32 & 1.3\\
27 &-18.35& 3.791$\times10^{-14}$ & 28.83&0.60 & $>10$Gyr& $>10$Gyr & 0.5 & 0.5 & 16.22\\
28 &-17.61&2.39$\times10^{-14}$ & 38.94&0.28 & 9Gyr & 25Myr & 0.88 & 0.12 & 0.02\\
29  &-18.00& 3.27$\times10^{-14}$ & 48.94&0.87 & 2Gyr& 1.3Gyr & 0.88 & 0.12 & 4.38\\
 \hline

\end{tabular}
\end{footnotesize}
\end{center}
Notes: Object no. 12 was shown to be a binary galaxy, with two BCG components separated by 48 kpc (Pustilnik et al. 2002), measured separately by us. Object no. 4 shows two H$\alpha$-emitting regions, but was measured as a whole. The models were calculated with respect to the total flux from each galaxy entry in the table.
    \end{minipage}
\end{table*}

\section{Discussion}
The current SFR and the simplified SFH of the sample galaxies are presented in Table \ref{models}. Most galaxies show present SFRs of $\sim0.6$ M$_{\odot}$ yr$^{-1}$ (median value), while a few show a higher SFR of several $M_{\odot}$ yr$^{-1}$. %The median SFR of the sample is 0.72 $M_{\odot}/yr$.
Typical BCG SFRs were examined before; usually, median SFRs are $\sim 10^{-1}$ M$_{\odot}$ yr$^{-1}$  up to several M$_{\odot}$ yr$^{-1}$. Kong (2004) summarized previous work and discussed differences caused by different SFR indicators. For example, the median SFR in his 72 heterogeneous BCG sample was $\sim$2.3 $M_{\odot}$ yr$^{-1}$, using an H$\alpha$ indicator. Kong used also other SFR indicators and found that not correcting for internal extinction and for the underlying stellar populations would underestimate the SFR. In order to compare the result obtained here to those of Kong we recalculated the median SFR of Kong's sample, excluding galaxies much brighter or much fainter than those present in our sample (i.e., excluding galaxies with M$_{B}<-20$ or M$_{B}>-17$ $mags$). This yields a median SFR of $\sim$1 $M_{\odot}$ yr$^{-1}$, which is only slightly higher than ours probably due to the fact that we did not correct for internal extinction.

A more interesting result is obtained when examining the galaxy stellar compositions and SF histories. SFHs of BCG have also been examined before and a common belief is that old stellar populations, of one to a few Gyr, are present along the newly formed stars. The SF in such galaxies is believed to take place in short bursts separated by long quiescent periods (Kong 2003, Kong et al. 2003). Such a behaviour, of series of short SF bursts, was simulated by e.g. Valcke et al. (2008) for isolated dwarf galaxies. However, we find that the observed colours are better fitted when the stellar population is explained  by a combination of a continuous, constant SFR and a more recent instantaneous burst. As Table \ref{models} shows, the calculated burst times correspond to these assumptions, since most galaxies had their first SF process beginning several Gyr ago, whereas almost all galaxies show in addition a much recent burst.

Table \ref{models} shows that the simple-minded algorithm used here to evaluate the SFHs of the sample galaxies does sometimes produce unexpected results. An example is the model result for galaxies no. 7, 14, 22 and 27. There, no recent burst is fit by the models, and the galaxy colours are well explained by a constant SFR, which started $\sim10$Gyr ago. Moreover, for all the sample galaxies the observed light is dominated ($\sim80$\%) by the constant SFR, and only a small fraction (usually up to $\sim20$\%) is contributed by the recent burst.

We mentioned above the claim of Pustilnik et al.\ (2001a) that in at least $80\%$ of BCGs  the SF could have been triggered by tidal interactions or mergers, whereas about $43\%$ of the isolated BCGs are probably tidally influenced by dwarf galaxies in their vicinity. We did not identify obvious interactors that could have triggered SF in the ``isolated'' galaxies of our sample. This emphasizes that interactions may not be the primary triggers of SF in isolated galaxies. We checked this by plotting the SFR from column 4 in Table~4 against the number of nearest neighbours N$_{NN}$ from column 6 of Table~1. There is no correlation between the two (R$^2$=0.04), which shows that a slight increase of the galaxy density in the neighbourhood of an object does not significantly encourage the SF. Essentially, we find no SFR differences between objects in extremely underdense environments, such as the two components of SBS1120+586, and e.g.,  G0957+3503 the galaxy in the densest environment of our sample that has 15 NED-listed neighbours within 3 Mpc and 300 km sec$^{-1}$. %However, the inspection of SDSS images and the comments on individual objects given in the previous section shows a large fraction of cases where potential immediate neighbours, not recognized as separate galaxies, could be present.

Another interesting comparison is with the SFRs of late-type dwarf galaxies (LTDGs) in the Virgo cluster (Almoznino \& Brosch 1998a, b; Heller et al. 1999). The LTDGs are galaxies devoid of large-scale internal SF triggers such as spiral density waves. They are a legitimate comparison sample to the BCGs studied here, since these objects were measured in the same way as the galaxies studied here, with the same telescope and filters, and their SFRs were calculated with exactly the same method as used here. As the comparison sample consists of galaxies that are members of the Virgo Cluster, their N$_{NN}$ values are at least two orders of magnitude higher than for the BCGs studied here. However, since the initial selection specified low surface brightness, it is to be expected that the LTDGs would show also low SFR values compared with those of the BCG sample studied here.

We find that the BCGs studied here have SFRs higher by two-three orders of magnitude than those of the Virgo late-type dwarfs, despite their being relatively isolated. Part of the difference can be attributed to the intrinsic brightness of the galaxies. The BCGs are also significantly brighter than the BCDs and LSB Irr's studied by Almoznino \& Brosch (1998a, b) and by Heller et al. (1999), although about half of the objects included in the present sample do qualify as dwarfs (M$_B\geq$--18 mag). These are plotted in Figure 1, which shows M$_B$ from column 2 of Table 4 against log(SFR). The Virgo dwarfs were plotted in Figure 1 by assuming that all are at the same distance as the cluster, 18 Mpc, as adopted in our previous papers.

The correlation between the total starlight and H$\alpha$ is well known. Gavazzi \& Scodeggio (1996) found that galaxy colours interpreted as the age of the stellar population correlate with the H-band luminosity; the H-band luminosity was assumed to measure the stellar mass of a galaxy while bluer colours indicate younger stellar populations.  Frick et al. (2001) found a similar relation for HII regions in the nearby spiral galaxy NGC 6946 and Heller et al. (1999) showed this for dwarf galaxies in the Virgo cluster. Kong (2004) found that the SFR correlated with M$_B$ for the BCGs in his sample. The relation he found:
\begin{equation}
SFR=-(7.65 \pm 0.25)-(0.42 \pm 0.01)M_B
\end{equation}
fits also the points plotted by us, although the low-SFR Virgo dwarfs deviate below the relation. The Virgo dwarfs extend Kong's relation by two orders of magnitude, down to SFR$\simeq10^{-4}$ M$_{\odot}$ yr$^{-1}$. The BCGs studied here form more stars {\bf because} they have more stars to begin with, and their extreme isolated location does not influence significantly their SF properties when compared to dwarf galaxies in the Virgo cluster. This property, that the SFR correlates with the stellar light (or the stellar mass) is known also for high-redshift galaxy samples (e.g., Noeske et al. 2007; Elbaz et al. 2007; Daddi et al. 2007).

We showed above that, although the BCGs studied here are not absolutely isolated in that some may have a few neighbours within 300 km sec$^{-1}$ and 3 Mpc, they reside in regions of very low galaxy densities. We found that all are forming stars at the present time and that their SFRs are higher than those of Virgo Cluster dwarfs, but not exceedingly higher.
Being both compact and relatively isolated, no external triggers seem to have encouraged SF in these galaxies and the primary SF trigger is probably an internal one,  such as accretion of gas retained in their DM halos.
This trigger could be encouraged by tidal interactions, but is evidently active even in the absence of such interactions.

\section{Summary}
We studied the star formation properties and histories of galaxies in a sample of BCG galaxies located in voids or in void walls, selected to be as isolated as possible, in order to minimize the chance that external effects influenced the formation and evolution of these objects. BCGs are known to have an older population underlying the recently formed stars, usually explained as formed by several instantaneous SF bursts separated by long quiescent periods. Our stellar population analysis confirms the presence of a significant older population. We found that for practically all the galaxies a combination of a  constant SFR combined with a   recent instantaneous SF burst is a better explanation of the integrated stellar population.
The measured SFR is correlated with the existing stellar population, as found for other galaxies. %The SFRs we measured are in accord with other studies of BCGs.

%{\bf MORE HERE}

%% If you wish to include an acknowledgments section in your paper,
%% separate it off from the body of the text using the \acknowledgments
%% command.

%% Included in this acknowledgments section are examples of the
%% AASTeX hypertext markup commands. Use \url without the optional [HREF]
%% argument when you want to print the url directly in the text. Otherwise,
%% use either \url or \anchor, with the HREF as the first argument and the
%% text to be printed in the second.

\section{Acknowledgments}

We thank the anonymous referee for significant comments that improved the paper. We are grateful to Ulrich Lindner for producing the original sample on which this study was based, and to Ms. Hava Zilka for help with the data reduction effort.
This publication makes use of results of the Sloan Digital Sky Survey (SDSS) at http://www.sdss.org/. Funding for the SDSS and
SDSS-II has been provided by the Alfred P. Sloan Foundation, the
Participating Institutions, the National Science Foundation, the
U.S. Department of Energy, the National Aeronautics and Space
Administration, the Japanese Monbukagakusho, the Max Planck
Society, and the Higher Education Funding Council for England.

The SDSS is managed by the Astrophysical Research Consortium for
the Participating Institutions. The Participating Institutions are
the American Museum of Natural History, Astrophysical Institute
Potsdam, University of Basel, Cambridge University, Case Western
Reserve University, University of Chicago, Drexel University,
Fermilab, the Institute for Advanced Study, the Japan
Participation Group, Johns Hopkins University, the Joint Institute
for Nuclear Astrophysics, the Kavli Institute for Particle
Astrophysics and Cosmology, the Korean Scientist Group, the
Chinese Academy of Sciences (LAMOST), Los Alamos National
Laboratory, the Max-Planck-Institute for Astronomy (MPA), the
Max-Planck-Institute for Astrophysics (MPIA), New Mexico State
University, Ohio State University, University of Pittsburgh,
University of Portsmouth, Princeton University, the United States
Naval Observatory, and the University of Washington.

This publication makes use of data products from the Two Micron
All Sky Survey, which is a joint project of the University of
Massachusetts and the Infrared Processing and Analysis
Center/California Institute of Technology, funded by the National
Aeronautics and Space Administration and the National Science
Foundation.

%%%%%%%%%%%%%%%%%%%%%%%%%%%%%%%%%%%%%%%%%%%%%%%%%%%%%%%%%%%%%%%%%%%%%%%%%%%
\begin{figure*}
%\label{fig:SFR_MB}
{\centering
 \includegraphics[clip=,angle=-90,width=15.0cm]{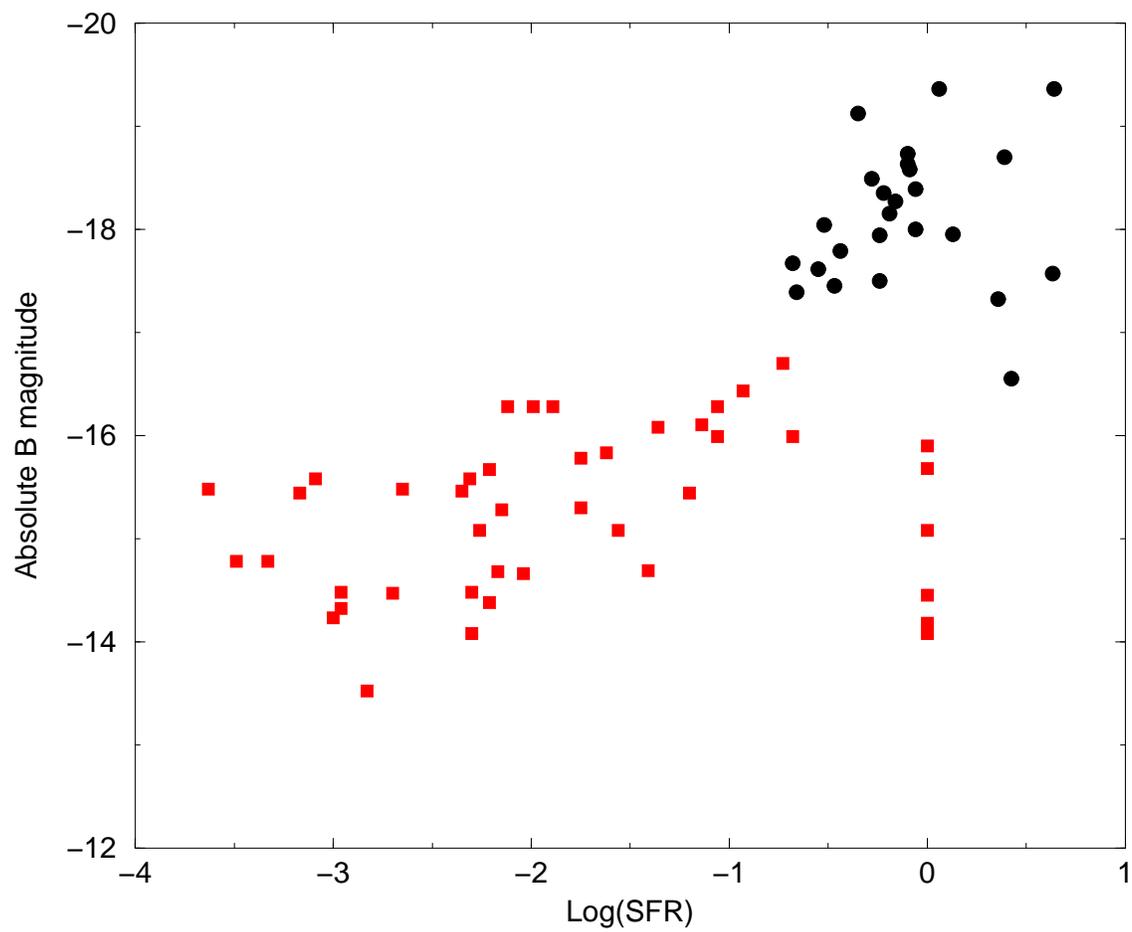}
}
 \caption{Blue absolute magnitude vs. SFR  for Virgo dwarf galaxies (red squares) and for the BCGs studied here (black circles).}
\end{figure*}
%%%%%%%%%%%%%%%%%%%%%%%%%%%%%%%%%%%%%%%%%%%%%%%%%%%%%%%%%%%%%%%%%%%%%%%%%%%

\end{document}